\documentclass[11pt]{article}
\usepackage{epsfig}
\usepackage{rotating}

\newcommand{\BABARPubYear}    {08}
\newcommand{\BABARConfNumber} {010}
\newcommand{\SLACPubNumber} {13335}

\input babarsym

\long\def\inst#1{\par\nobreak\kern 4pt\nobreak
    {\it #1}\par\vskip 10pt plus 3pt minus 3pt}

\begin{document}
{\pagestyle{empty}

\begin{flushright}
\babar-CONF-\BABARPubYear/\BABARConfNumber \\
SLAC-PUB-\SLACPubNumber \\
\end{flushright}

\par\vskip 4cm

\begin{center}
\Large \bf Evidence for {\boldmath $B$} semileptonic decays into the charmed baryon
  {\boldmath $\Lambda_c^+$} 
\end{center}
\bigskip

\begin{center}
\large The \babar\ Collaboration\\
\mbox{ }\\
\today
\end{center}
\bigskip \bigskip

\begin{center}
\large \bf Abstract
\end{center}
We present the first evidence for $B$ semileptonic decays into the charmed
baryon $\Lambda_c^+$ based on 420~fb$^{-1}$ of data collected at the
$\Upsilon(4S)$ resonance with the \babar\ detector at the \pep2\ $e^+e^-$
storage rings.  Events are tagged by fully reconstructing one of the $B$ mesons
in a hadronic decay mode. We measure the relative branching fraction
${\cal B}(\overline{B}
  \to \Lambda^+_c X \ell^- \overline{\nu}_{\ell})/{\cal B}(\overline{B} \to
  \Lambda^+_c/\overline{\Lambda}^-_c X)=
    (3.2\pm0.9_{\rm stat.}\pm0.9_{\rm syst.})\%$.
The significance of the signal including the systematic uncertainty
is 4.9 standard deviations.
\vfill
\begin{center}

Submitted to the 34$^{\rm th}$ International Conference on High-Energy Physics, 
ICHEP 08,\\
29 July---5 August 2008, Philadelphia, Pennsylvania.

\end{center}

\vspace{1.0cm}
\begin{center}
{\em Stanford Linear Accelerator Center, Stanford University,
Stanford, CA 94309} \\ \vspace{0.1cm}\hrule\vspace{0.1cm}
Work supported in part by Department of Energy contract DE-AC03-76SF00515.
\end{center}

\newpage
}

\begin{center}
\small

The \babar\ Collaboration,
\bigskip

%
B.~Aubert,
M.~Bona,
Y.~Karyotakis,
J.~P.~Lees,
V.~Poireau,
E.~Prencipe,
X.~Prudent,
V.~Tisserand
\inst{Laboratoire de Physique des Particules, IN2P3/CNRS et Universit\'e de Savoie, F-74941 Annecy-Le-Vieux, France }
J.~Garra~Tico,
E.~Grauges
\inst{Universitat de Barcelona, Facultat de Fisica, Departament ECM, E-08028 Barcelona, Spain }
% ...

\end{center}\newpage

\section{Introduction}
\label{sec:Introduction}

\begin{figure}[hbtp]
\begin{center}
\epsfig{figure=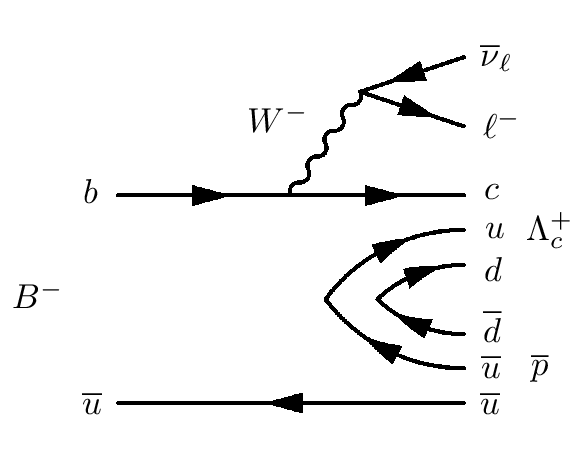}
\epsfig{figure=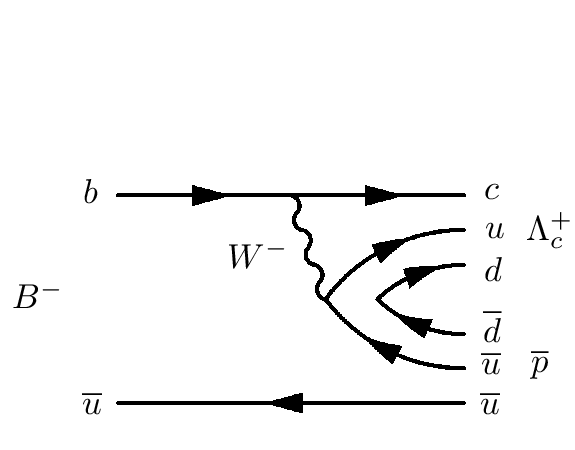}
\end{center}
\caption{ $B$ decays with an external $W$ emission (left)
  and an internal $W$ emission (right).}
\label{fig_diagram}
\end{figure}

The $B$ decays to charmed baryons are not as well understood as the decays into
charmed mesons.  In particular, there is limited knowledge, both theoretical
and experimental, about the $B$ semileptonic decays into charmed baryons.  If
the charmed baryonic production is dominated by the emission of an external $W$
boson, as in the case of the mesonic production $B\rightarrow DX$, then we can
expect the rate of the semileptonic events to be about the same for the
baryonic and mesonic processes:
\begin{equation}
\frac{{\mathcal B}(\overline{B} \to D X \ell^- \overline{\nu}_{\ell})}
     {{\mathcal B}(\overline{B} \to D/\overline{D} X)}
\sim
\frac{{\mathcal B}(\overline{B}\to \Lambda^+_c X \ell^- \overline{\nu}_{\ell})}
     {{\mathcal B}(\overline{B}\to  \Lambda^+_c/\overline\Lambda_c^- X)}
\end{equation}
where $\ell = e$ or $\mu$.
For the mesonic process, the ratio is
${\mathcal B}(\overline{B} \to D X \ell^- \overline{\nu}_{\ell})/
 {\mathcal B}(\overline{B} \to D/\overline{D} X)
\approx 10\%$~\cite{pdg}.
If the baryonic ratio is found to be smaller than the mesonic ratio,
there is a significant contribution from internal $W$ emission
in the baryonic production.
The Feynman diagrams of $B$ decays with internal and external
$W$ emissions are shown in Fig.~\ref{fig_diagram}.

About 90\% of the inclusive semileptonic
$\overline{B} \to X \ell^- \overline{\nu}_{\ell}$
branching fraction~\cite{babarExcl} can be accounted for by
summing the branching fractions from exclusive
$\overline{B} \rightarrow D^{(*)}(\pi) \ell^- \overline{\nu}_{\ell}$ decays.
Semileptonic $\overline{B}$ decays to 
charmed baryons may account for some of the remaining difference.
These decays have not yet been observed.  A previous search for
$\overline{B} \to  \Lambda^+_c X e^-\overline{\nu}_e$ by the CLEO
collaboration \cite{CLEO} obtained an upper limit of
${\cal B}(\overline{B} \to \Lambda^+_c X e^-\overline{\nu}_e)/
 {\cal B}(\overline{B} \to  \Lambda^+_c / \overline{\Lambda}^-_c X)<0.05$
 at the 90\% confidence level.

In this paper, we present the first evidence for
$\overline{B}\to \Lambda_c^+ X e^- \overline{\nu}_{e}$
decays\footnote{Charge-conjugate modes are implied
throughout this paper.}, where $X$ can be
any particle(s) from the $B$ semileptonic decay other than
the leptons and the $\Lambda_c^+$.
The $\overline{B} \to \Lambda_c^+ X e^- \overline{\nu}_{e}$ signal yield is
obtained by a fit to the $\Lambda^+_c$ invariant mass. We perform a blind
analysis using Monte-Carlo (MC) samples and the $\Lambda^+_c$ invariant mass sidebands on data
to optimize the selection criteria and estimate the backgrounds.  We also present
the results for a similar search for
$\overline{B} \to \Lambda_c^+ X \mu^- \overline{\nu}_{\mu}$.

\section{The \babar\ Detector and Dataset}
\label{sec:babar}
This analysis is based on data collected with the \babar\ detector at the
\pep2\ storage rings. The total integrated luminosity of the dataset is
420~fb$^{-1}$ collected on the $\Upsilon(4S)$ resonance. The corresponding number
of produced \BB\ pairs is roughly 460 million.
An additional 42 fb$^{-1}$
data sample taken at a center-of-mass (CM) energy 40 MeV below the $\Upsilon(4S)$
resonance is used to study background from $e^+e^- \to
f\overline{f}~(f=u,d,s,c,\tau)$ events (continuum production).
The \babar\ detector
is described in detail elsewhere~\cite{detector}. Charged-particle trajectories
are measured by a 5-layer double-sided silicon vertex tracker and a 40-layer
drift chamber, both operating in a 1.5-T magnetic field. Charged-particle
identification is provided by the average energy loss (d$E$/d$x$) in the
tracking devices and by an internally reflecting ring-imaging Cherenkov
detector. Photons are detected by a CsI(Tl) electromagnetic calorimeter.
Muons are identified by the instrumented magnetic-flux return. 
A detailed GEANT4-based MC simulation~\cite{Geant} of \BB\ and
continuum events has been used to study the detector response, its acceptance,
and to test the analysis techniques. 

\section{Simulation of
  {\boldmath $\overline{B} \to \Lambda_c^+ X \ell^- \overline{\nu}_{\ell}$} Decays}  
\label{sec:signalMC}

Due to a lack of available theoretical models for semileptonic
$B$ decays to charmed baryons, we use an ad-hoc model to study
selection optimization in our analysis.  In our 
model, the $B$ decays semileptonically into charmed baryons
through an intermediate massive particle $Y$ as $\overline{B} \to Y
\ell^-\overline{\nu}_{\ell}$, according to a phase space model~\cite{jetset}.
The $Y$ subsequently decays into a $\Lambda^+_c$,
an anti-nucleon (anti-proton or anti-neutron),
and $n_1$ ($n_2$) charged (neutral) pions, with $n_1$ and $n_2$
distributed as Poissonians with a common mean of $1.25$.
We apply isospin symmetry in the final states of the $Y$ decay.

\begin{figure}[!ht]
\centering
\epsfig{figure=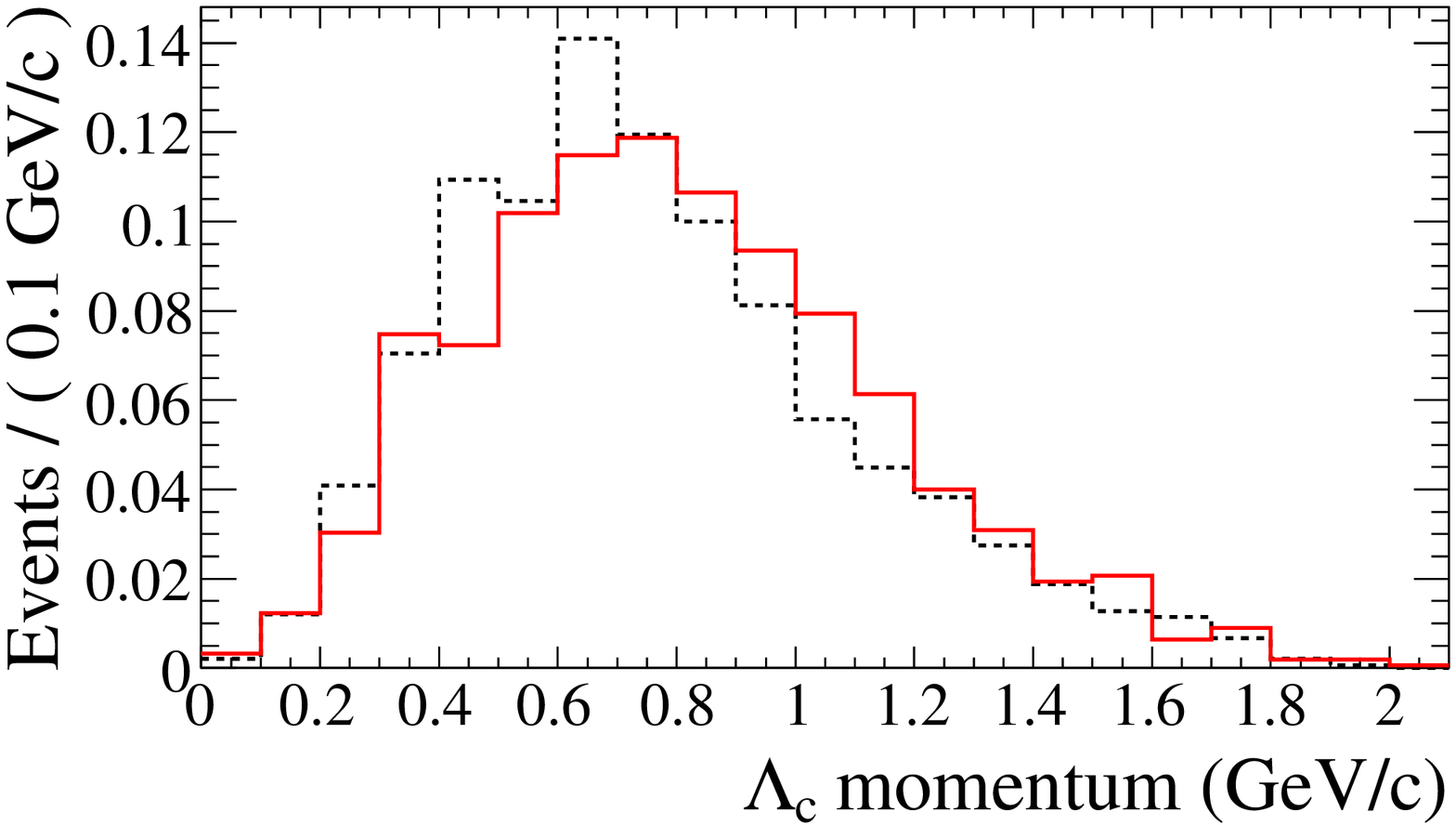,width=8.1cm}
\epsfig{figure=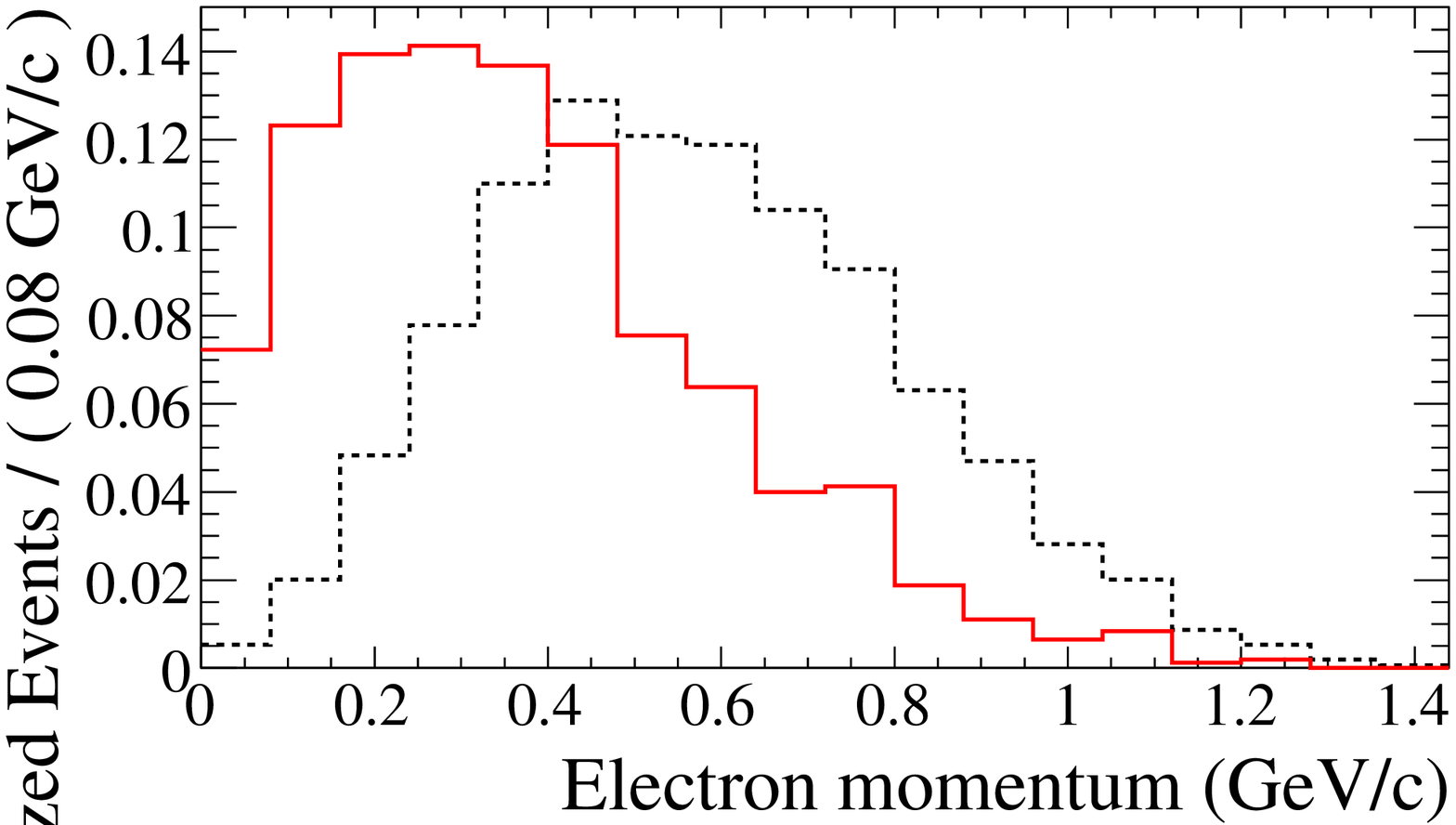,width=8.1cm}
\caption{$\Lambda^+_c$ and lepton momentum spectra for a sample of signal MC
  events using the models before tuning (dashed line)
  and after tuning (solid line) as described in the text.}
\label{fig:Lclepmom}
\end{figure}

Due to the ad-hoc nature of the signal model, we tune its parameters
as part of the analysis.
We perform a first optimization of the selection criteria using
a pseudo-particle $Y$ mass $m_Y $ of 4.0
GeV/$c^2$ and a width $\Gamma_Y$ of 0.6 GeV/$c^2$, 
with the $Y$ decay constrained by $n_1 + n_2 \leq 4$.
After unblinding the $\Lambda^+_c$ invariant mass signal region, we compare the
sideband-subtracted signal distributions
of the $\Lambda^+_c$ and lepton momentum spectra, and the charged
and neutral pion multiplicity, with the corresponding ones from the signal MC
simulation.  We then tune the signal model parameters to resemble the observed
distributions on data.
For instance, by adjusting the mass and width of the pseudo-particle $Y$, we can
control the shape of the lepton momentum spectrum.
We find a good agreement with the electron spectrum in data for $m_Y=4.5$
GeV/$c^2$, $\Gamma_Y=0.2$ GeV/$c^2$, and $n_1 +n_2 \leq 6$.
We use the retuned signal model to re-optimize the selection criteria and to
estimate signal efficiency.
We check
{\em a posteriori} that the two-pass selection criteria optimization
does not strongly depend on the initial signal model parameters used.
The $\Lambda^+_c$ and lepton momentum spectra for a sample of signal MC events
are shown in Fig. \ref{fig:Lclepmom} for two the models before and after tuning.

\section{Event Selection}
\label{sec:Analysis}
We select $B$ semileptonic decays in events containing a fully
reconstructed $B$ meson ($B_{\rm tag}$), which allows us to constrain the
kinematics, to reduce the combinatorial background, and to determine the charge
and flavor of the signal $B$, up to flavor-changing mixing.

We first reconstruct the $B$ semileptonic decay, selecting a lepton with
momentum $p^*_{\ell}$ in the CM frame higher than 0.35
GeV/$c$. Electrons from photon conversion and $\pi^0$ Dalitz decay are removed
using a dedicated algorithm, which uses a least-squares fit to reconstruct the
vertex between two tracks of opposite charges whose kinematical parameters are
compatible with a photon conversion or a $\pi^0$ Dalitz decay.  Tracks identified
as electrons or
muons are required to be within 0.1 cm of the
the interaction point at their point of closest approach in the transverse plane.
Candidate $\Lambda^+_c$ baryons, with the correct charge-flavor correlation with
the lepton, are reconstructed in the $p K^-\pi^+$, $p K^0_S$, $p K^0_S \pi^+
\pi^-$, $\Lambda \pi^+$, $\Lambda \pi^+ \pi^+ \pi^-$ modes.
$K^0_S$ ($\Lambda$) candidates are reconstructed in the $\pi^+\pi^-$ ($p\pi^-$)
decay mode.

In events with multiple $\overline{B} \to \Lambda^+_c X \ell^-
\overline{\nu}_{\ell}$ candidates, the candidate with the highest
$\Lambda^+_c$-$\ell^-$ vertex fit probability is selected.

We reconstruct $B_{\rm tag}$ decays of the type $\overline{B} \rightarrow D Y'$,
where
$Y'$ represents a collection of hadrons with a total charge of $\pm 1$, composed
of $n_1'\pi^{\pm}+n_2' K^{\pm}+n_3' K^0_S+n_4'\pi^0$,
   where $n_1'+n_2' \leq  5$, $n_3' \leq 2$,
   and $n_4' \leq 2$. Using $D^0$ $(D^+)$ and $D^{*0}$ $(D^{*+})$ as seeds for
$B^-$ $(\overline{B^0})$ decays, we reconstruct about 1000 types of decay chains.
For each of the $B_{\rm tag}$ decay modes, the purity $\mathcal P$ is
estimated using MC simulation. $\mathcal P$ is defined as the ratio of signal
over background events with $m_{\rm ES}\ge 5.27$ GeV/$c^2$.

The kinematic consistency of a $B_{\rm tag}$ candidate
is checked using two variables: the beam-energy substituted mass
$m_{ES}=\sqrt{s/4-\vec{p}_B^2}$, and the energy difference $\Delta E = E_B
-\sqrt{s}/2$. Here $\sqrt{s}$ refers to the total CM  energy, while $\vec{p}_B$
and $E_B$ denote the momentum and energy of the $B_{\rm tag}$ candidate in the
CM frame. For correctly identified $B_{\rm tag}$ decays, the $m_{ES}$
distribution peaks at the $B$ mass, while $\Delta E$ peaks at
zero.  We select a $B_{\rm tag}$ candidate in the signal region defined as
5.27~GeV/$c^2$ $< m_{ES} <$ 5.29~GeV/$c^2$, excluding $B_{\rm tag}$ candidates
with daughter particles in common with the charmed baryon or the lepton from the
$B$ semileptonic decay.

In the case of multiple $B_{\rm tag}$ candidates, we select the one with the
largest purity $\mathcal P$ of the $B_{\rm tag}$ mode; in the case of
multiple candidates with the same $B_{\rm tag}$ mode
we select the one with the smallest $|\Delta E|$ value.

The $B_{\rm tag}$,
$\Lambda^+_c$, and $\ell^-$ candidates are required to have the correct
charge-flavor correlation.

By fully reconstructing one $B$ in the event and by requiring the presence of a
proton or $\Lambda$ candidate, used for the $\Lambda^+_c$ reconstruction, the
resulting sample shows a high purity, as is generally the case for a tagged
analysis.  In order to minimize model-dependent
effects that can be introduced by a specific set of selection criteria, we only
require the lepton momentum to be greater than 0.35 GeV$/c$, as described
above, and the missing momentum to be greater  than 0.2 GeV$/c$. This last cut
removes background from hadronic $\overline{B} \to \Lambda^+_c X$ decays in
which all the particles in the $X$ system have been reconstructed and one
hadron is misidentified as a lepton.
We also require the total charge of the reconstructed event
to be zero, in order to reduce combinatorial background in the $B_{\rm tag}$
reconstruction from missing particles. 

To obtain the $B$ semileptonic signal yields, we perform a one-dimensional
binned maximum likelihood fit to the $\Lambda^+_c$ invariant mass distribution. 
Backgrounds to the process of interest can be divided according to whether they contain
a correctly-reconstructed $\Lambda^+_c$ with a mass value in the signal region
(peaking), and those that do not.
MC studies of generic \BB\ and continuum events show that the peaking background
comes mainly from hadronic $\overline{B}\to \Lambda^+_c X$ decays,
with a fake electron from gamma conversions or $\pi^0$ Dalitz decays, or a hadron
misidentified as a muon.
The number of peaking background events from hadronic $\overline{B}\to \Lambda^+_c X$
decays is estimated from the simulation, as:
\begin{displaymath}
N_{\rm peak} = 2 \cdot \epsilon_{\rm peak} \cdot {\cal B} (\overline{B} \to \Lambda^+_c X) \cdot N_{\BB}
\end{displaymath}
where $\epsilon_{\rm peak}$ is the efficiency of reconstructing fake
$\overline{B}\to \Lambda^+_c X \ell^- \overline{\nu}_{\ell}$ events in a hadronic
$\overline{B}\to \Lambda^+_c X$ sample
computed as the ratio of reconstructed and generated events,
and $N_{\BB}$ is the number of \BB\ pairs
corresponding to the data luminosity.
We use
${\cal B} (\overline{B} \to \Lambda^+_c X)=(3.9 \pm 1.2) \%$,
which is the statistical average of the charged and neutral $B$
branching fractions~\cite{pdg}.
We estimate
$N_{\rm peak} =  5.2\pm 1.0_{\rm stat.}\pm 1.7_{\rm syst.}$ events and
$N_{\rm peak} = 15.3\pm 1.4_{\rm stat.}\pm 4.9_{\rm syst.}$ events for the
electron and muon samples, respectively.
The sources of systematic error are described in Sec.\ \ref{sec:Systematics}.

The $\Lambda^+_c$ invariant mass distribution is fitted with the sum of three
probability density functions (PDFs):
one Gaussian for semileptonic
$\overline{B} \rightarrow \Lambda^+_c X \ell^- \overline{\nu}_{\ell}$
events,
another Gaussian for peaking background events from hadronic $\overline{B}
\rightarrow \Lambda^+_c X$ decays, and a first order polynomial for
combinatorial \BB\ and continuum background events.
The two Gaussians share the same mean and width, and the 
width is fixed to the value obtained by fitting
the hadronic $\overline{B}\to\Lambda_c^+X$ data sample, as described in
Sec.\ \ref{sec:BFMeasurement}.
The amount of peaking background PDF
is fixed to the MC prediction. We first fit the $\Lambda^+_c$ invariant mass
sidebands, defined as the mass window of $2.23-2.26$, $2.31-2.34$ GeV/$c^2$,
to constrain the background PDF parameters. We then fit the
$\Lambda^+_c$ invariant mass including the signal region, where the free
parameters are the signal and background yields, and the $\Lambda^+_c$ mass
mean value. 

\section{Measurement of Branching Fractions}
\label{sec:BFMeasurement}

\begin{figure}[!ht]
\centering
\epsfig{figure=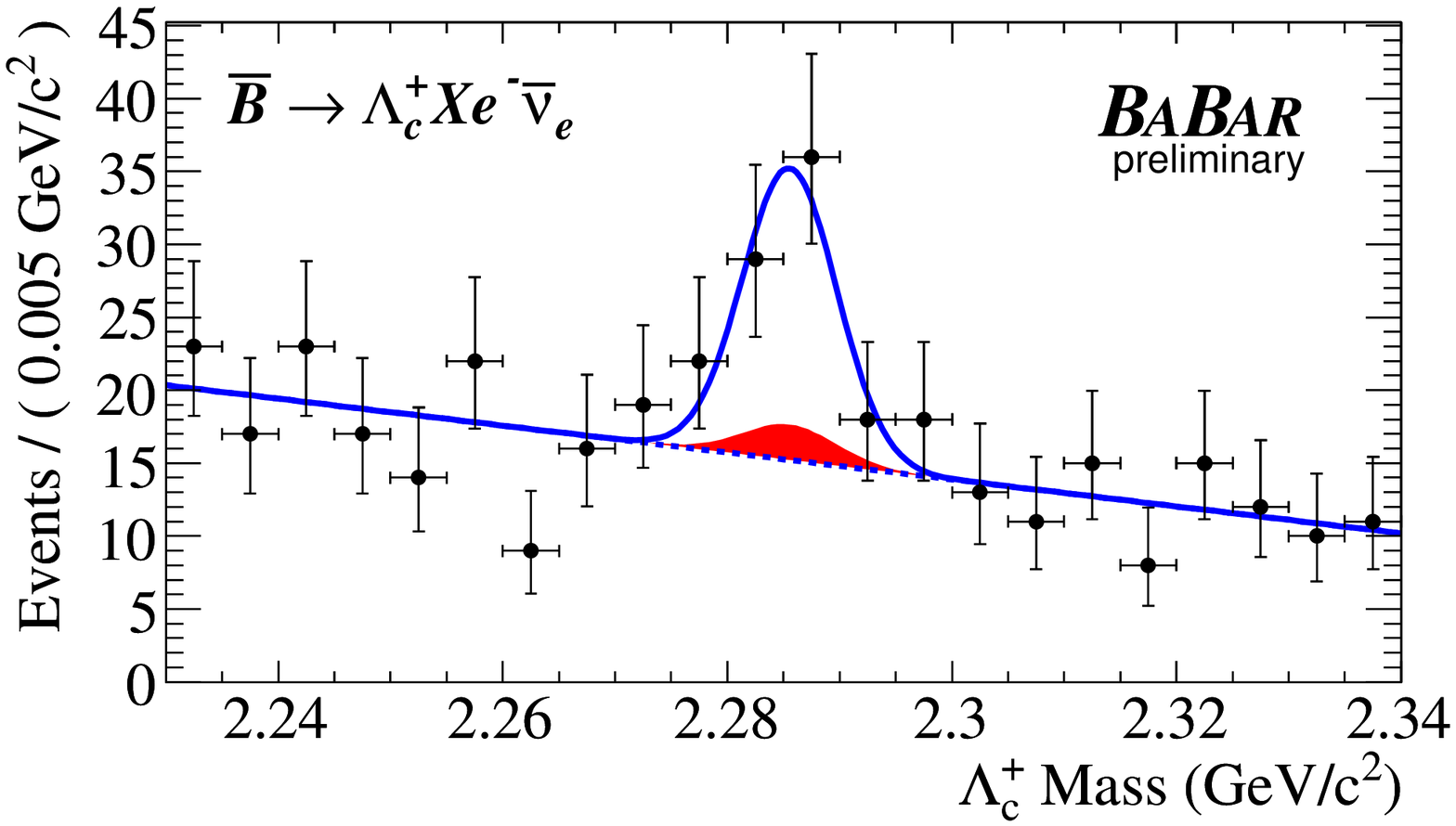,width=8.1cm}
\epsfig{figure=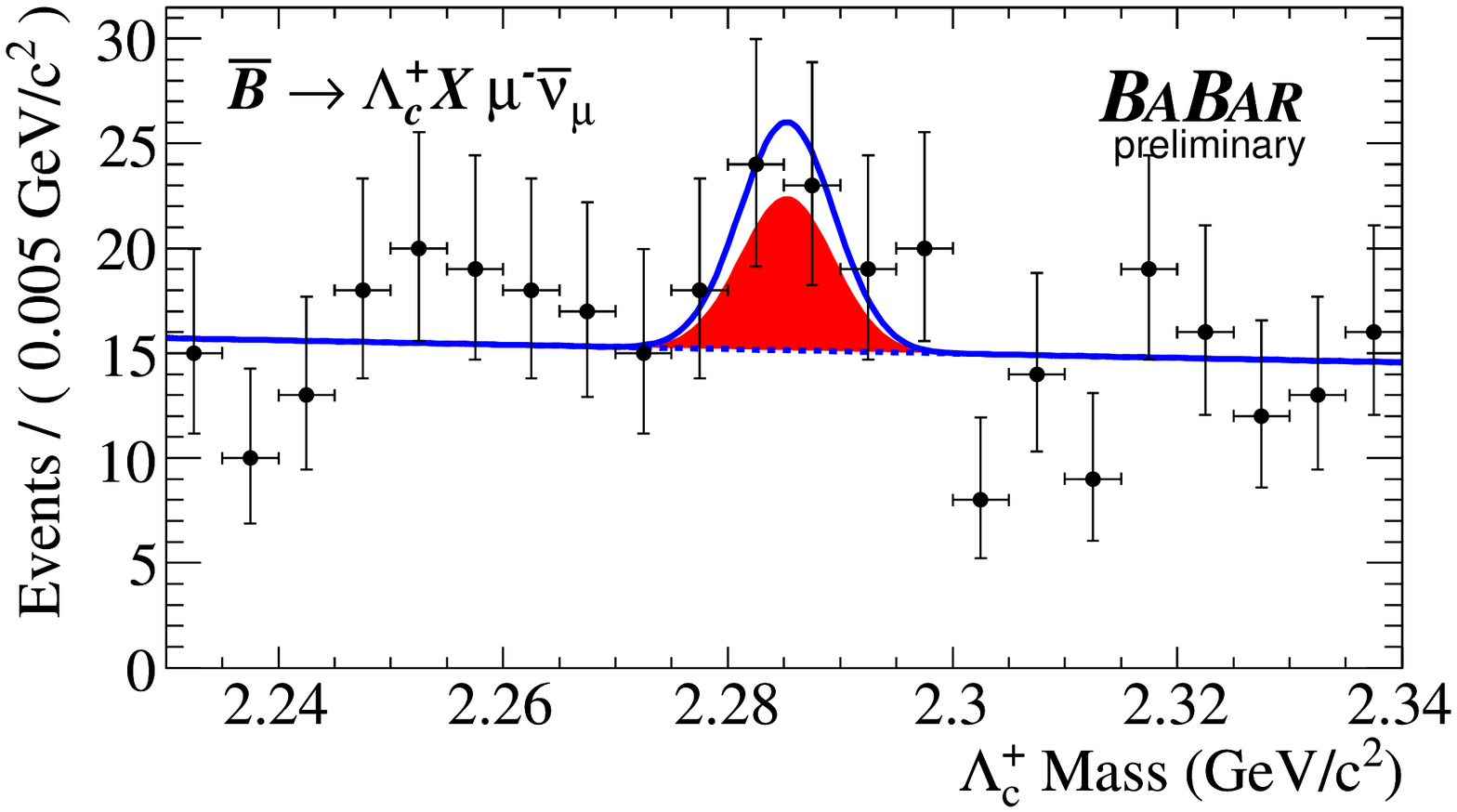,width=8.1cm}
\caption{Fit to the $\Lambda^+_c$ distribution for
$\overline{B} \to \Lambda^+_c X e^- \overline{\nu}_{e}$ (left) and
$\overline{B} \to \Lambda^+_c X \mu^- \overline{\nu}_{\mu}$ (right):
the data (points with error bars) are compared to the results of the overall
fit (solid line). The peaking background contribution is shown with a shaded area.  The combinatorial \BB\ and continuum background is shown with a dashed line.}
\label{fig:Fit1}
\end{figure}

The $\Lambda^+_c$ invariant mass distribution is compared with the results of
the fit in Figure \ref{fig:Fit1} for the $\overline{B} \to \Lambda^+_c X \ell^-
\overline{\nu}_{\ell}$ decays, where we show separately the electron and muon
sample.

\begin{figure}[!ht]
\centering
\epsfig{figure=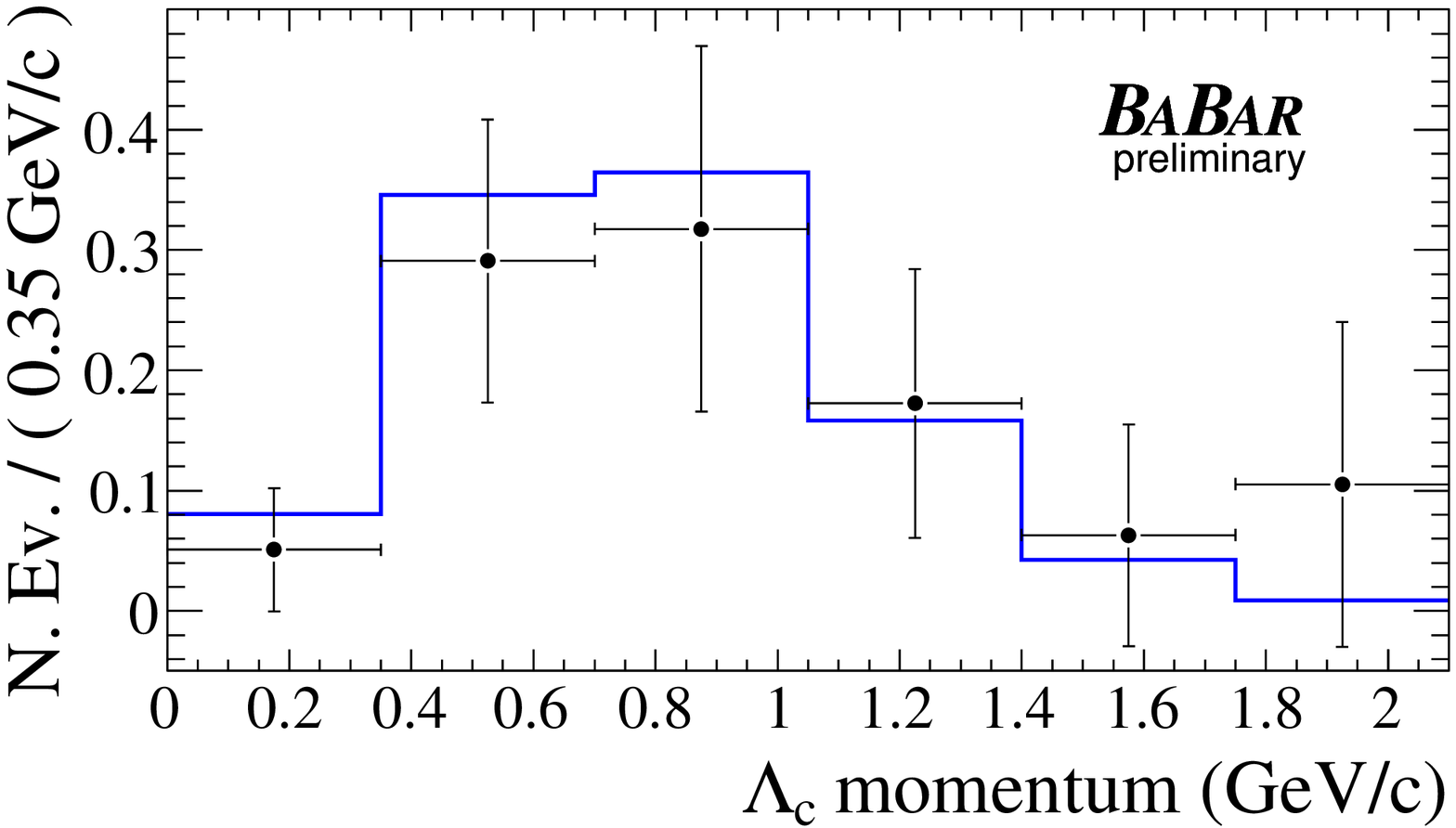,width=8.1cm}
\epsfig{figure=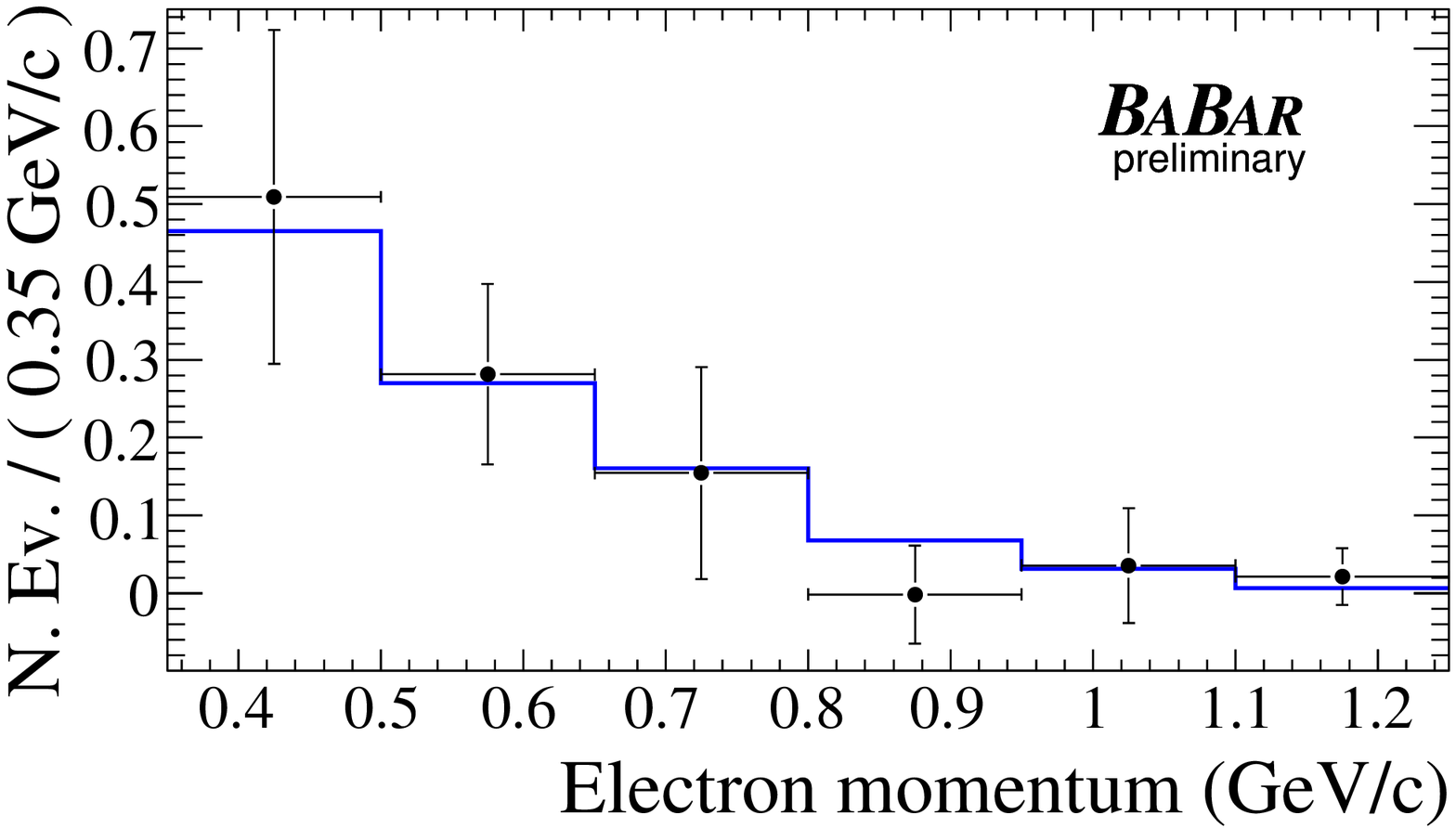,width=8.1cm}
\caption{Sideband-subtracted, efficiency-corrected, normalized distributions
for the $\Lambda^+_c$ and electron momentum spectrum in
$\overline{B}\to \Lambda^+_c X e^- \overline{\nu}_{e}$ decays:
the data (points with error bars) are compared to the signal MC prediction.}
\label{fig:SB-1}
\end{figure}

\begin{figure}[!ht]
\centering
\epsfig{figure=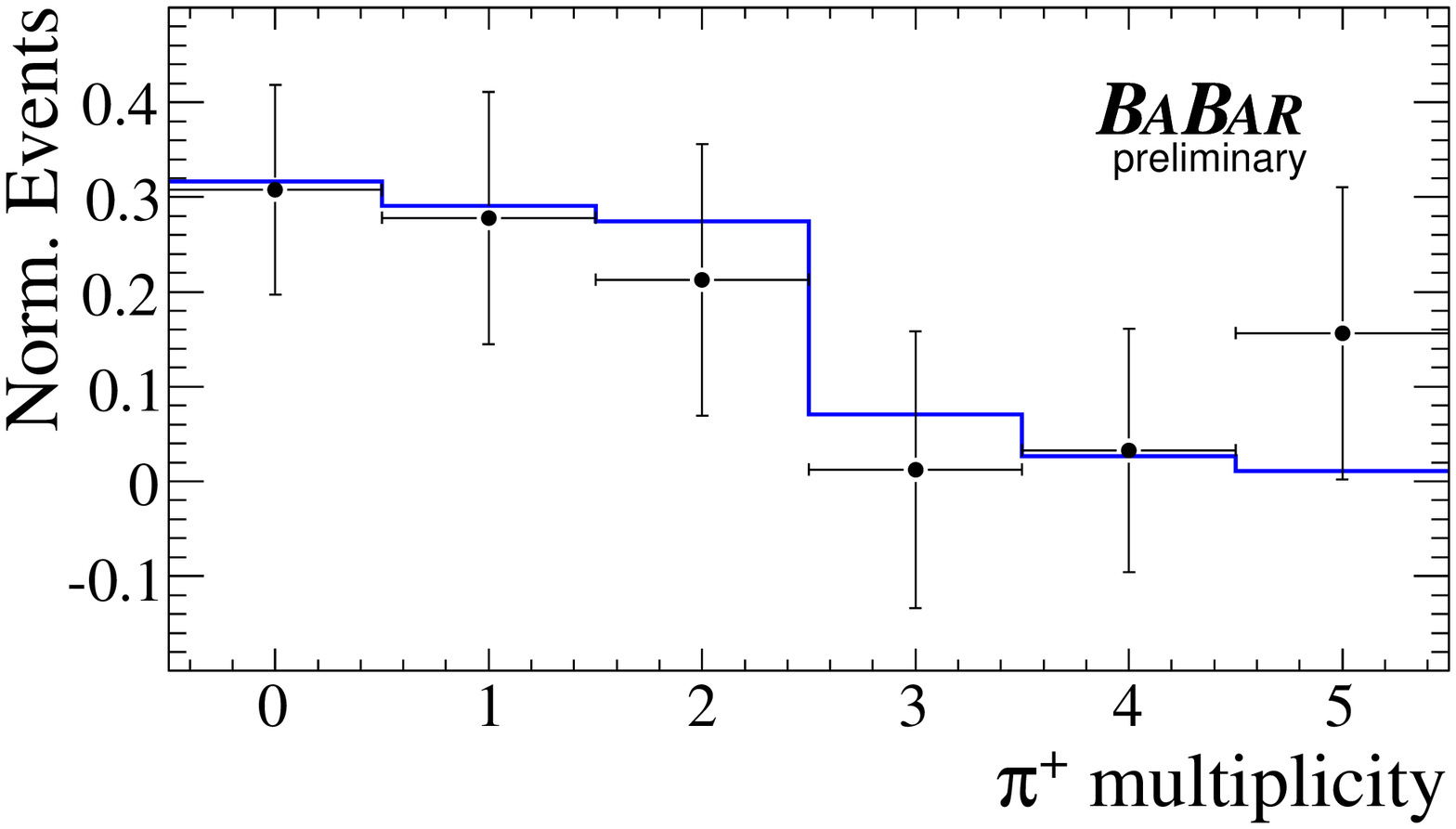,width=8.1cm}
\epsfig{figure=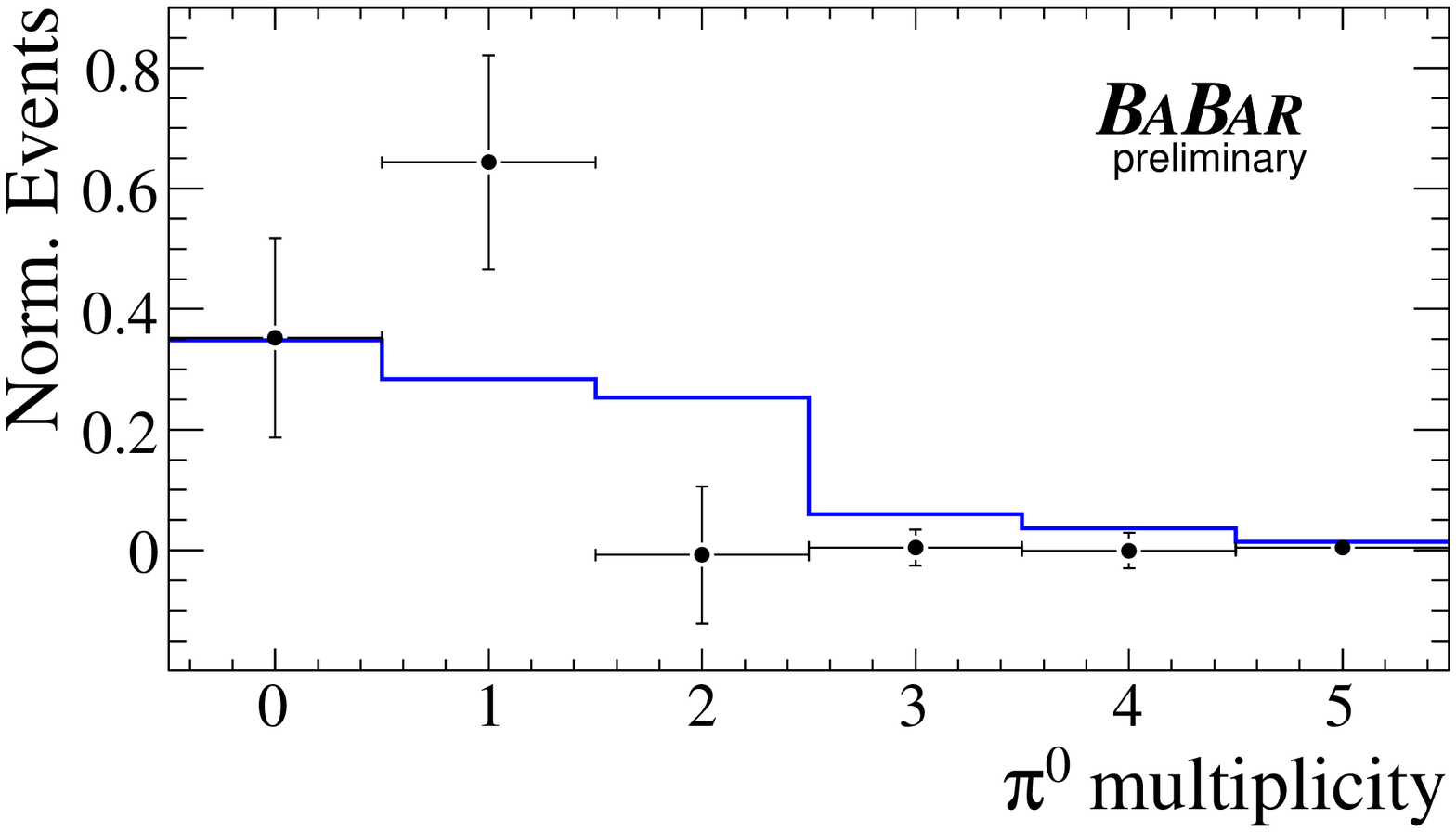,width=8.1cm}
\caption{Sideband-subtracted, efficiency-corrected, normalized distributions
for the charged and neutral pion multiplicity in the $X$ system in
$\overline{B}\to \Lambda^+_c X e^- \overline{\nu}_{e}$ decays:
the data (points with error bars) are compared to the signal MC prediction.}
\label{fig:SB-2}
\end{figure}

In Fig. \ref{fig:SB-1} and \ref{fig:SB-2}, we show the distributions for the
$\Lambda^+_c$ and electron momentum spectrum, and the charged and neutral pion
multiplicity in the $X$ system for the
$\overline{B}\to \Lambda^+_c X e^- \overline{\nu}_{e}$ sample.
These distributions are sideband-subtracted, by
selecting events in the $\Lambda^+_c$ invariant mass signal region, and
subtracting the combinatorial \BB\ and continuum background using the invariant
mass sidebands, and then corrected bin-by-bin by the
efficiency estimated from the signal MC.
The peaking background contribution is subtracted using the MC
prediction. 

In order to reduce the systematic uncertainty, the exclusive ${\cal B}
(\overline{B} \rightarrow \Lambda^+_c X \ell^- \overline{\nu}_{\ell})$ branching
fractions  are measured relative to the hadronic
${\cal B} (\overline{B} \rightarrow \Lambda^+_c/\overline{\Lambda}^-_c X )$ branching
fraction.  To determine the hadronic branching fraction,  we use
a sample of ${\cal B} (\overline{B} \rightarrow \Lambda^+_c/\overline{\Lambda}^-_c X)$
events selected similarly to the semileptonic channel.  We select the
reconstructed $\Lambda^+_c$ candidate with the highest vertex probability, and
a $B_{\rm tag}$ candidate in the signal region defined as
5.27~GeV/$c^2$ $< m_{ES} <$ 5.29~GeV/$c^2$, excluding $B_{\rm tag}$ candidates
with daughter particles in common with the charmed baryon.
In the case of multiple $B_{\rm tag}$ candidates, we select the one with the
largest {\em a priori} purity of the $B_{\rm tag}$ mode; in the case of
multiple candidates with the same $B_{\rm tag}$ mode (same {\em a priori}
purity), we select the one with the smallest $|\Delta E|$ value.

\begin{figure}[!ht]
\centering
\epsfig{figure=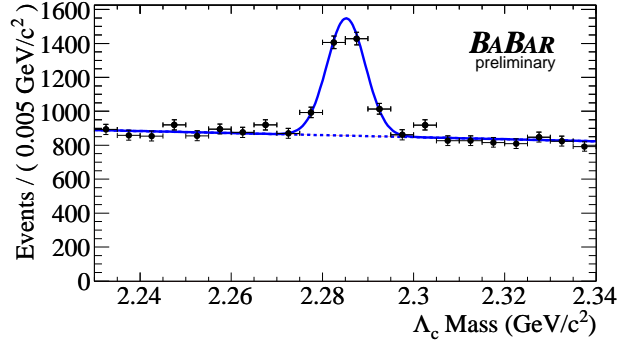,width=8.1cm}
\caption{Fit to the $\Lambda^+_c$ distribution for
$\overline{B} \to \Lambda^+_c / \overline{\Lambda}^-_c X$: the data (points with error bars)
are compared to the results of the overall fit (solid line).
The combinatorial \BB\ and continuum background is shown with a dashed line.}
\label{fig:Fit2}
\end{figure}

To obtain the $\overline{B} \to \Lambda^+_c/\overline{\Lambda}^-_c X$ signal yield, we
perform a one-dimensional binned maximum likelihood fit to the $\Lambda^+_c$
invariant mass distribution with the sum of two PDFs: a single Gaussian
for hadronic $\overline{B} \to \Lambda^+_c/\overline{\Lambda}^-_c X$ events,
and a first order polynomial background for combinatorial \BB\ and continuum
background events. The $\Lambda^+_c$ invariant mass distribution is compared
with the results of the fit in Figure \ref{fig:Fit2}.

\begin{table}[!htb]
\centering
\caption{Signal yields and reconstruction efficiencies for the
$\overline{B} \to \Lambda^+_c X \ell^- \overline{\nu}_{\ell}$ and
$\overline{B} \to \Lambda^+_c /\overline{\Lambda}^-_c X$ decays
  with statistical errors.}
\begin{tabular}{lcc}
\hline
\hline
Decay Mode & $N_{\rm data}$ & $\epsilon$ $(\times 10^{-5})$  \\
\hline
$\overline{B} \rightarrow \Lambda^+_c X e^- \overline{\nu}_{e}$
& $38 \pm 10$
& $2.06\pm 0.17$ \\
$\overline{B} \rightarrow \Lambda^+_c X \mu^- \overline{\nu}_{\mu}$
& $7.4 \pm 8.4$
& $1.07\pm 0.12$ \\
$\overline{B} \rightarrow \Lambda^+_c/\overline{\Lambda}^-_c X$
& $1432 \pm 81$
& $3.02\pm 0.13$ \\
\hline
\hline
\end{tabular}
\label{tab:results}
\end{table}

The relative branching fraction
${\cal B} (\overline{B} \to \Lambda^+_c X\ell^- \overline{\nu}_{\ell})/
{\cal B} (\overline{B} \to \Lambda^+_c/\overline{\Lambda}^-_c X)$ is obtained by
correcting the signal yields obtained from the fit by the reconstruction
efficiency ratio:
\begin{displaymath}
\frac{{\cal B}(B\to\Lambda^+_cX\ell^- \overline{\nu}_{\ell})}{{\cal B}(B\to\Lambda^+_c/\overline{\Lambda}^-_cX)} =
\left(\frac{N_{\rm semil}}{N_{\rm had}}\right)
\left(\frac{\epsilon_{\rm had}}
{\epsilon_{\rm semil}}\right).
\end{displaymath}
Here, $N_{\rm semil}$  ($N_{\rm had}$) is the number of $\overline{B} \to \Lambda^+_c X$
signal events for the semileptonic~(hadronic) mode, reported in
Table~\ref{tab:results} together with the corresponding reconstruction
efficiencies $\epsilon$.

\section{Systematic Uncertainties}
\label{sec:Systematics}
By measuring the $\overline{B} \to \Lambda^+_c X\ell^- \overline{\nu}_{\ell}$
branching fraction relative to the
$\overline{B} \to \Lambda^+_c/\overline{\Lambda}^-_c X$ branching fraction,
many uncertainties in the reconstruction efficiency ratio cancel out.
In particular, the uncertainties in the
$\Lambda^+_c$ and $B_{\rm tag}$ reconstruction efficiencies and the
$\Lambda^+_c$ decay branching fractions do not contribute.

We categorize the remaining systematic uncertainties into additive
uncertainties which directly affect the signal yield, and multiplicative
uncertainties which affect only the branching fraction ratio.
The systematic uncertainties that have been considered are described below and
summarized in Tab.~\ref{tab:r24-systematic}.

\begin{table}[ht]
\centering
\caption{Table of systematic uncertainties.
The additive errors (errors on the signal yield)
are shown in units of events.
The multiplicative errors (errors on the reconstruction efficiency)
are shown in units of percent.
}
\label{tab:r24-systematic}
\begin{tabular}{lcc}
\hline\hline
Additive systematics &
$\overline{B}\to\Lambda_c^+Xe^-\overline{\nu}_e$ &
$\overline{B}\to\Lambda_c^+X\mu^-\overline{\nu}_\mu$ \\
\hline
Peaking background statistics (ev.)
  & $1.0$  & $1.4$ \\
Uncertainty in ${\mathcal B}(\overline{B}\to\Lambda_c^+X)$ (ev.)
  & $1.6$  & $4.7$ \\
Lepton misidentification rate (ev.)
  & $0.7$  & $2.0$ \\
Fit bias (ev.)
  & $0.3$  & $1.2$ \\
\hline
Total additive (ev.)
  & $2.0$  & $5.4$ \\
\hline\hline
Multiplicative systematics &
$\overline{B}\to\Lambda_c^+Xe^-\overline{\nu}_e$ &
$\overline{B}\to\Lambda_c^+X\mu^-\overline{\nu}_\mu$ \\
\hline
Reconstruction efficiency statistics (\%)
  & $8.4$ & $11.4$ \\
Model dependence (lepton momentum) (\%)
  & $22.6$ & $71.8$ \\
Model dependence ($\Lambda_c^+$ momentum) (\%)
  & $7.5$ & $7.5$ \\
Lepton identification efficiency (\%)
  & $1.1$  & $2.7$ \\
Selection order (\%)
  & $3.7$ & $44.6$ \\
\hline
Total multiplicative (\%)
  & $25.5$ & $85.7$ \\
\hline\hline
\end{tabular}
\end{table}

\subsection{Additive Systematics}

Systematic uncertainties in the signal yield are dominated by the
peaking-background yield estimate.
We evaluate this uncertainty by propagating
the uncertainty in the $\overline{B}\to\Lambda^+_c X$ branching fraction,
and the Poisson error due to the limited hadronic MC statistics.
We also vary the lepton
misidentification probabilities by 15\% for both electrons and muons, and
add in quadrature the corresponding variation in the peaking background rate.
To evaluate the effect of a bias in the fit technique, we perform
ensembles of MC experiments, in which events are generated according to
the PDF shapes measured on data, varying the signal to background rate,
and fitting for the signal as in the full analysis.  The difference between
the fitted value of the yield and the true value is taken as a systematic error.

\subsection{Multiplicative Systematics}

Systematic uncertainties which affect the reconstruction efficiency ratio
are dominated by the uncertainty in the signal model.
To estimate this, we look for deviations in the
reconstruction efficiency as we vary the tuning parameters of the signal model.
The deviation in the electron spectrum is taken as the
$\Delta\chi^2/{\rm n.d.f.}=1$ difference, where
$\Delta\chi^2$ is the data-MC difference divided by the statistical error
on data, added in quadrature for each bin of momentum, and ${\rm n.d.f.}=6$
is the number of bins.
The deviation in the $\Lambda_c^+$ spectrum is taken as
the variation in the efficiency as a function of $\Lambda_c^+$ momentum.
We also include the Poisson error contribution of the limited signal MC
statistics.
We estimate the systematic uncertainty due to particle identification by
varying the electron (muon) identification efficiency by 2\% (3\%).
Because the event selection order is slightly different in the semileptonic and
hadronic sample selections, the reconstruction efficiency systematics do not
exactly cancel.
We evaluate the corresponding systematic uncertainty by reversing the order
of the lepton and $B_{\rm tag}$ selection, taking the difference in the
signal yield, corrected by the reconstruction efficiency, as the systematic
error.  The large systematic uncertainty coming from the
selection order in the muon channel
is due to the low statistics of the muon sample.

\section{Results}
\label{sec:results}

We measure the following branching ratios:
\begin{eqnarray}
\frac{{\cal B}(B\to\Lambda^+_cXe^- \overline{\nu}_{e})}
     {{\cal B}(B\to\Lambda^+_c/\overline{\Lambda}^-_cX)}&=&
     (3.9\pm1.0_{\rm stat.}\pm1.1_{\rm syst.})\% \\
\frac{{\cal B}(B\to\Lambda^+_cX\mu^- \overline{\nu}_{\mu})}
     {{\cal B}(B\to\Lambda^+_c/\overline{\Lambda}^-_cX)}&=&
     (1.5\pm1.7_{\rm stat.}\pm1.7_{\rm syst.})\%.
\end{eqnarray}
The result for the electron channel is compatible with the CLEO
result~\cite{CLEO} of
${\cal B}(B\to\Lambda^+_cXe^- \overline{\nu}_{e})$
/
${\cal B}(B\to\Lambda^+_c/\overline{\Lambda}^-_cX)<0.05$.
Despite the lower value of the muon result, the two channels are
compatible within their errors.
Taking the weighted average of the two channels, we obtain:
\begin{eqnarray}
\frac{{\cal B}(B\to\Lambda^+_cX\ell^- \overline{\nu}_{\ell})}
     {{\cal B}(B\to\Lambda^+_c/\overline{\Lambda}^-_cX)}&=&
     (3.2\pm0.9_{\rm stat.}\pm0.9_{\rm syst.})\%.
\end{eqnarray}

By using the hadronic branching fraction
${\cal B}(B\to\Lambda^+_c/\overline{\Lambda}^-_c X)=(4.5\pm1.2)\%$~\cite{pdg},
we obtain: 
\begin{eqnarray}
{\cal B}(B\to\Lambda^+_cXe^- \overline{\nu}_{e})&=&
    (1.8\pm0.5_{\rm stat.}\pm0.7_{\rm syst.})\times10^{-3} \\
{\cal B}(B\to\Lambda^+_cX\mu^- \overline{\nu}_{\mu})&=&
    (6.6\pm7.5_{\rm stat.}\pm7.6_{\rm syst.})\times10^{-4}\\
{\cal B}(B\to\Lambda^+_cX\ell^- \overline{\nu}_{\ell})&=&
    (1.5\pm0.4_{\rm stat.}\pm0.6_{\rm syst.})\times10^{-3}.
\end{eqnarray}

We evaluate the significance ${\mathcal S}$
of our result using the log likelihood ratio
$-2\log({\mathcal L}_{\rm max}/{\mathcal L}_0)$ where
${\mathcal L}_{\rm max}$ is the maximum likelihood and
${\mathcal L_0}$ is the likelihood value fixing the signal yield to be zero,
including statistical and systematic errors:
\begin{eqnarray}
{\mathcal S}_e =
\sqrt{-2\log({\mathcal L}_{\rm max}/{\mathcal L}_0)} &=& 4.6 \\
{\mathcal S}_\mu =
\sqrt{-2\log({\mathcal L}_{\rm max}/{\mathcal L}_0)} &=& 0.8
\end{eqnarray}
where the subscript $e$ and $\mu$ denotes the electron and muon channels,
respectively.
The probability $P$ that our measured signal results from a background
fluctuation are computed from these ${\mathcal S}$ values to be
$P_e=2.2\times10^{-6}$ and
$P_\mu=0.22$.
In order to compute the significance for the combined result,
we take the product of these two probabilities:
$P_\ell=4.8\times10^{-7}$.  Converting this back to the significance,
we obtain ${\mathcal S}_\ell=4.9$ for the combined result.

In conclusion, we present the first evidence for $B$ semileptonic decays into
the charmed baryon $\Lambda^+_c$. The relative branching fraction
${\cal B}(B\to\Lambda^+_cX\ell^- \overline{\nu}_{\ell})/
{\cal B}(B\to\Lambda^+_c/\overline{\Lambda}^-_cX)$
is found to be smaller than the corresponding one for the $D$ charmed meson. We
measure the $\Lambda^+_c$ and electron momentum spectrum in the $B$ semileptonic
decay, which could be of guidance to further development in the theoretical
modeling of these decays. 

\section{Acknowledgments}
\label{sec:Acknowledgments}

We are grateful for the 
extraordinary contributions of our \pep2\ colleagues in
achieving the excellent luminosity and machine conditions
that have made this work possible.
The success of this project also relies critically on the 
expertise and dedication of the computing organizations that 
support \babar.
The collaborating institutions wish to thank 
SLAC for its support and the kind hospitality extended to them. 
This work is supported by the
US Department of Energy
and National Science Foundation, the
Natural Sciences and Engineering Research Council (Canada),
the Commissariat \`a l'Energie Atomique and
Institut National de Physique Nucl\'eaire et de Physique des Particules
(France), the
Bundesministerium f\"ur Bildung und Forschung and
Deutsche Forschungsgemeinschaft
(Germany), the
Istituto Nazionale di Fisica Nucleare (Italy),
the Foundation for Fundamental Research on Matter (The Netherlands),
the Research Council of Norway, the
Ministry of Education and Science of the Russian Federation, 
Ministerio de Educaci\'on y Ciencia (Spain), and the
Science and Technology Facilities Council (United Kingdom).
Individuals have received support from 
the Marie-Curie IEF program (European Union) and
the A. P. Sloan Foundation.

\end{document}